\documentclass{ws-mpla}
\usepackage[super]{cite}
\usepackage{graphicx}
\usepackage{color}
\usepackage[]{hyperref}
\begin{document}

\markboth{Y.Li and L.Y.Zhang}{Inflationary magnetogenesis of primordial magnetic fields with multiple vector fields}

\catchline{}{}{}{}{}

\title{Inflationary magnetogenesis of primordial magnetic fields with multiple vector fields
}

\author{Yu Li\footnote{
Corresponding author.}}

\address{School of Science, Dalian Maritime University, Dalian 116026, China\\
leeyu@dlmu.edu.cn}

\author{Le-Yao Zhang}

\address{School of Science, Dalian Maritime University, Dalian 116026, China\\
	zhangleyao@dlmu.edu.cn}

\maketitle

\pub{Received (Day Month Year)}{Revised (Day Month Year)}

\begin{abstract}
In this paper, we discussed the multiple vector fields during the inflation era and the inflationary magnetogenesis with multiple vector fields. Instead of a single coupling function in single vector field models, the coupling matrix between vector fields and scalar field which drive the inflation is introduced. The dynamical equations for multiple vector fields are obtained and applied to the inflation era. We discussed three cases for the double-field model. In no mutual-coupling case, one can find that both electric and magnetic spectrum can be scale-invariant at the end of inflation, meanwhile, the strong coupling problem can be avoided. The effect of mutual- coupling between different vector fields is also discussed. We found that weak mutual-coupling can lead to a slightly blue spectrum of the magnetic field. On the other hand, in the strong mutual-coupling case, the scale-invariant magnetic spectrum can also be obtained but the energy density of electromagnetic fields either leads to the backreaction problem or is diluted by inflation.

\end{abstract}

\ccode{PACS Nos.:98.80.Cq.}

\section{Introduction}
Observations show that our universe is magnetized on a wide range of length scales 
\cite{Kronberg_1994,RevModPhys.74.775,10.1007/978-94-017-3239-0_21,Clarke_2001,doi:10.1126/science.1184192,stevenson2010planetary,taylor2011extragalactic}.
The origin of these magnetic fields remains unclear. One class of schemes to explain the origin of these magnetic fields is the astrophysical scenario in which the magnetic fields originate from some astrophysical processes \cite{10.1093/mnras/271.1.L15,Kulsrud_1997,Gnedin_2000,rees2006origin}. These schemes can explain the origin of magnetic fields in galaxies and clusters. However, this type of model is difficult to explain the origin of the magnetic fields in cosmic voids. These large-scale magnetic fields seem more likely to have originated in the early universe\cite{doi:10.1126/science.1184192,archambault2017search,ackermann2018search}.

Another class of schemes is the primordial scenario\cite{GRASSO2001163,durrer2013cosmological,kandus2011primordial,subramanian2016origin}, in which these large-scale magnetic fields are assumed to have originated in the early stages of the universe.
One class of possible sources of primordial magnetic fields is phase transitions like an electroweak phase transition\cite{vachaspati1991magnetic,sigl1997primordial,vachaspati2021progress} or the QCD transition\cite{kisslinger2003magnetic,tevzadze2012magnetic}. 
However, in these scenarios, very tiny fields on galactic scales are obtained unless helicity is also generated in which case one can have an inverse cascade of energy to large scales\cite{brandenburg1996large,banerjee2004evolution}.

The other class of possible sources of the primordial magnetic field is inflationary magnetogenesis \cite{turner1988inflation,ratra1992cosmological,martin2008generation,Kobayashi_2014,PhysRevD.89.063002,campanelli2015lorentz,Tasinato_2015,PhysRevD.96.083511,PhysRevD.97.083503,PhysRevD.100.023524,Fujita_2019,li2022inflationary}. Although inflation provides an ideal setup for large-scale field generation\cite{turner1988inflation}, several problems remain in the mechanism of inflationary magnetogenesis.

Firstly, the electromagnetic field is not amplified during the inflation era because of the conformal invariance of the standard electromagnetic action and the Friedmann-Robertson-Walker (FRW) metric is conformally flat\cite{PhysRevLett.21.562}. Therefore, one necessary condition for the generation of large-scale primordial magnetic fields during inflation is the breaking of the conformal invariance of the electromagnetic field action\cite{turner1988inflation,ratra1992cosmological,martin2008generation,PhysRevD.48.2499,PhysRevLett.75.3796,PhysRevD.62.123505,fujita2015consistent,subramanian2010magnetic}. One method to do this is to introduce a time-dependent coupling function $f^2(\phi)$ into the action \cite{ratra1992cosmological}. By this method, the scale-invariance spectrum of the magnetic field can be obtained \cite{PhysRevD.96.083511,PhysRevD.97.083503}. 

Secondly, once the coupling function was introduced, the effective charge can be defined by $e/f$ \cite{Demozzi_2009}. If one requires an effective charge consistent with Maxwell's theory at the end of inflation, then $f$ should tend to $1$ at the end of inflation. If $f$ is an increasing function of time, then $f$ will be very small at the beginning of inflation, which means that the effective charge is very huge at the beginning of inflation. This would imply that the interaction between charged particles and the electromagnetic fields would be extremely strong. This is the so-called strong coupling problem. To avoid the strong coupling problem, $f$ should be a decreasing function.  

Thirdly, however, if $f$ decreases from a large value to $1$, the electric energy density will increas rapidly during the inflation era and it would eventually exceed the inflation energy density \cite{PhysRevD.96.083511}. This problem is known as the backreaction problem. The backreaction problem will constrain the energy density of inflation and lead to the Schwinger effect during inflation which can further stop the generation of the magnetic field \cite{Kobayashi_2014_a}. 

In \cite{PhysRevD.96.083511}, the authors introduced a transition at the end of inflation in the evolution of $f$ to avoid the strong coupling problem. However, their model still constrains the inflation energy density and reheating temperature if it is required to avoid the backreaction problem.

Strong coupling problem and backreaction problem cannot be avoided at the same time if only one coupling function is introduced since a function cannot be an increasing function and a decreasing function at the same time. Therefore a possible solution is to introduce more than one coupling function with different evolutionary behaviors. One method to do this is to consider the multiple vector fields models.

The multiple vector fields models during the inflation era are widely discussed. However, most of the works focus on the isotropic vector inflation scenario\cite{golovnev2008vector,maleknejad2013gauge,PhysRevD.84.043515,PhysRevLett.102.111301,PhysRevD.79.063517,PhysRevD.80.123530,Gorji_2020}.
The reason why multiple vector fields are considered in these works is to ensure the isotropy of the space-time background. However, in this paper, we treat the multiple vector fields as perturbations and they would not affect the evolution of the inflation which is driven by a scalar field.

This paper is organized as follows: we derive the Hamiltonian equations for multiple vector fields in curved spacetime in Sec.\ref{s2}. The power spectrum of electromagnetic fields during the inflation era is obtained in Sec.\ref{s3}. In Sec.\ref{s4}, we discuss three cases of the double-field model. The summary and discussion are in Sec.\ref{s5}.

\section{Multiple vector fields in curved spacetime\label{s2}}
In this section we are going to derive the dynamical equations of multiple vector fields by the Hamiltonian method starting with the following action:
\begin{equation}
	\label{e1}
	S=\int d^4x \sqrt{-g}(\mathcal{L}_g+\mathcal{L}_{\phi}+\mathcal{L}_{EM})
\end{equation}
where the gravitational field, scalar field, and vector fields parts are respectively:
\begin{align}
	\mathcal{L}_g&=\frac{1}{16\pi}R\label{e2}\\
	\mathcal{L}_{\phi}&=-\frac{1}{2}g^{ab}\nabla_a\phi\nabla_b\phi-V(\phi)\label{e3}\\
	\mathcal{L}_{EM}&=-\frac{1}{16\pi}\left[G_{AB}(\phi)g^{ac}g^{bd}F_{cd}^AF_{ab}^B+\tilde{G}_{AB}(\phi)g^{ac}g^{bd}\tilde{F}_{cd}^AF_{ab}^B\right]
	\label{e4}
\end{align}
in which $a,b,\dots$ are abstract spacetime indexes introduced by R. Penrose \cite{penrose1984spinors}, $A,B,\dots$ are multiple vector fields indexes to denote different vector fields. 

The electromagnetic field tensor $F^A_{ab}$ is defined as:
\begin{equation}
	F^A_{ab}:=\nabla_aA_b^A-\nabla_bA_a^A=\partial_aA_b^A-\partial_bA_a^A\label{e5}
\end{equation}
where $A^A_a$ are vector fields. The dual form of $F^A_{ab}$ is defined as:
\begin{equation}
	\tilde{F}^A_{ab}:=\frac{1}{2}(F^{cd})^A\epsilon_{cdab}\label{e6}
\end{equation}
where $\epsilon_{cdab}$ is the volume element which is compatible with the metric of spacetime $g_{ab}$,
{and in any (right-handed) coordinate basis it takes the form \cite{wald2010general}:
\begin{equation}
	\label{e76}
	\epsilon_{abcd}=\sqrt{|g|}(dx^0)_a\wedge(dx^1)_b\wedge(dx^2)_c\wedge(dx^3)_d
\end{equation}
}
$G_{AB}$ and $\tilde{G}_{AB}$ are symmetric coupling matrices whose matrix elements are functions of scalar field $\phi$.

To get the Hamiltonian, the spacetime should be 3+1 decomposed. Without loss of generality, we employ a non-standard 3+1 decomposition\footnote{\textcolor{black}{In \cite{liangGR}, "standard 3+1 decomposition" represents the orthogonal decomposition i.e. $N^a=0,N=1$.}}, in which the observer 4-velocity $Z^a$ is decomposed by
\begin{equation}
	\label{e7}
	Z^a=Nn^a+N^a
\end{equation}
where $N$ is the lapse function, $N^a$ is the shift vector field and $n^a$ is space hypersurface orthogonal timelike unit vector. 

\textcolor{black}{
The dynamic equations of $A_a^A$ can be obtained by demanding that the action \eqref{e1} is stationary under the variation of $A_a^A$. However, since we consider the coupling of the vector fields with the scalar field, the constraint algebra of the system will be different from ordinary electrodynamics. This issue arises in some inflationary magnetogenesis models e.g. \cite{li2022inflationary}.  Therefore, in this paper, we adopt the Hamiltonian approach to derive the dynamic and constraint equations.
}

To make sure that the electric and magnetic fields are spatial vector fields, they should be defined as 
\begin{equation}
	\label{e15}
	E_a^A:=n^bF_{ab}^A,~~~~~~~~~~B_a^A:=n^b\tilde{F}_{ba}^A
\end{equation}

The configuration variables of the electromagnetic field can be selected as the time component and spatial component of $A_a^A$:
\begin{equation}
	\mathbb{A}^A_0:=Z^aA_a^A,~~~~\mathbb{A}_a^A:=h^b_aA_b^A\label{e8}
\end{equation}
where $h^b_a:=\delta^b_a+n^bn_a$ is the spatial projection operator. In this paper, we denote the derivative operator which is compatible with spatial metric $h_{ab}$ as $D_a$, and the time derivative operator of spatial tensor fields as $D_0$. Then, the conjugate momentum of $\mathbb{A}^A_0,\mathbb{A}_a^A$ are respectively:
\begin{align}
	\Pi_A^0:=&\frac{\partial(\sqrt{-g}\mathcal{L}_{EM})}{\partial (D_0\mathbb{A}_0^A)}=0\label{e9}\\
	\Pi_A^a:=&\frac{\partial(\sqrt{-g}\mathcal{L}_{EM})}{\partial (D_0\mathbb{A}_a^A)}\nonumber\\
	=&-\frac{\sqrt{h}}{4\pi N}G_{AB}h^{ab}\left[(D_b\mathbb{A}_0^B-D_0\mathbb{A}_b^B)-N^c(D_b\mathbb{A}_c^B-D_c\mathbb{A}_b^B)\right]\nonumber\\
	&+\frac{\sqrt{h}}{8\pi}\tilde{G}_{AB}\hat{\epsilon}_{cbd}h^{mc}h^{nb}h^{ad}(D_m\mathbb{A}_n^B-D_n\mathbb{A}_m^B)\label{e10}\\
	=&-\frac{\sqrt{h}}{4\pi}\left[G_{AB}(E^a)^B-\tilde{G}_{AB}(B^a)^B\right]\label{e10-a}
\end{align}
where $\hat{\epsilon}_{abc}:=-\epsilon_{abcd}n^d$ is induced  volume element on space hypersurface.

Eq.\eqref{e9} means that there is no $D_0\mathbb{A}_0^A$ term in $\mathcal{L}_{EM}$ which means that the \eqref{e9} is the primary constraint. 
The Hamiltonian density can be obtained by Legendre transformation:
\begin{equation}
	\mathcal{H}_{EM}:=\Pi^a_AD_0\mathbb{A}_a^A-N\sqrt{h}\mathcal{L}_{EM}
	=N\sqrt{h}\mathcal{H}_0+N^a\mathcal{H}_a+\mathbb{A}_0^B\mathcal{G}_B
	\label{e11}
\end{equation}
where
\begin{align}
	\mathcal{H}_0:=&\frac{2\pi}{h}G^{CD}h_{ab}\Pi_C^a\Pi_D^b+\frac{1}{16\pi}\left(G_{AB}h^{ac}h^{bd}+\frac{1}{2}\tilde{G}_{AB}\epsilon^{abcd}\right)\mathbb{F}_{ab}^A\mathbb{F}_{cd}^B\nonumber\\
	&+\frac{1}{8\pi}\tilde{G}_{AB}\tilde{G}_{CD}G^{BD}h^{ab}\tilde{\mathbb{F}}_a^A\tilde{\mathbb{F}}_b^C-\frac{1}{\sqrt{h}}G^{AC}\tilde{G}_{BC}\Pi_A^a\tilde{\mathbb{F}}_a^B\label{e18}\\	
	\mathcal{H}_a:=&\Pi_A^b\mathbb{F}_{ba}^A\label{e19}\\
	\mathcal{G}_B:=&-D_a\Pi_B^a
	\label{e20}
\end{align}
in which
\begin{equation}
	\label{e12}
	\mathbb{F}_{ab}^A:=h^c_ah^d_bF_{cd}^A=D_a\mathbb{A}_b^A-D_b\mathbb{A}_a^A,~~~~\tilde{\mathbb{F}}_c^A:=\frac{1}{2}\hat{\epsilon}_{abc}h^{ma}h^{nb}\mathbb{F}_{mn}^A
\end{equation}

The evolution of primary constraint \eqref{e9} can be obtained as
\begin{equation}
	\label{e14}
	\dot{\Pi}_A^0=-\frac{\delta H_{EM}}{\delta \mathbb{A}_0^A}=-\mathcal{G}_A
\end{equation}
Then the  secondary constraint (Gauss constraint) $\dot{\Pi}_A^0\approx 0$  lead to $\mathcal{G}_A\approx 0$ i.e.
\begin{equation}
	\label{e21}
	D_a\left[G_{AB}(E^a)^B-\tilde{G}_{AB}(B^a)^B\right]=0
\end{equation}

{
It can be seen that the Gaussian constraints \eqref{e21} are different from those in ordinary electrodynamics, especially when the coupling matrix is a function of spatial coordinates.	If the electromagnetic field is still required to satisfy the ordinary Gaussian constraints, then Eq.\eqref{e21} puts forward additional requirements on the form of the coupling matrix. Similar issues are discussed in the \cite{li2022inflationary}.
In this paper, we focus on the uniform scalar field in the FRW spacetime, so Eq.\eqref{e21} is the same as the ordinary Gaussian constraint.
}

$\mathbb{A}_0^A$ as a free Lagrange multiplier can be chosen arbitrarily. Therefore we adopt the Coulomb gauge which is 
\begin{equation}
	\label{e22}
	\mathbb{A}_0^A=0,~~~~D^a\mathbb{A}_a^A=0
\end{equation}
The Hamiltonian equations under the Coulomb gauge are:
\begin{align}
	\dot{\mathbb{A}}_a^A=&\frac{\delta H_{EM}}{\delta \Pi_A^a}=\frac{4\pi N}{\sqrt{h}}G^{AB}h_{ab}\Pi_B^b-NG^{AC}\tilde{G}_{BC}\tilde{\mathbb{F}}_a^B-N^b\mathbb{F}_{ab}^A\label{e23}\\
	\dot{\Pi}_A^a=&-\frac{\delta H_{EM}}{\delta \mathbb{A}_a^A}\nonumber\\
	=&-\hat{\epsilon}_{bc}^{~~a}D^c(G^{BC}\tilde{G}_{AC}\Pi_B^b)N+\frac{\sqrt{h}}{4\pi}\hat{\epsilon}^{bca}D_c(\tilde{G}_{CB}\tilde{G}_{AD}G^{BD}\tilde{\mathbb{F}}_b^CN)
	+\frac{\sqrt{h}}{4\pi}D_c[G_{AB}(\mathbb{F}^{ca})^BN]\nonumber\\
	&-\frac{\sqrt{h}}{4\pi}D_c(\tilde{G}_{AB}\mathbb{F}_{db}^BN\epsilon^{abcd})
	-2D_b(N^{[a}\Pi_A^{b]})\label{e24}
\end{align}

\section{Power spectrum of the electromagnetic field during the inflation era\label{s3}}
From now on, we consider the FRW metric as follows
\begin{equation}
	\label{e25}
	ds^2=-dt^2+a^2(t)\delta_{ij}dx^idx^j=a^2(\eta)(-d\eta^2+\delta_{ij}dx^idx^j)
\end{equation}
where $t$ is cosmic time and $\eta$ is conformal time. We use conformal time in the following discussion which means that $N=a(\eta)$ and $N^a=0$. Because the metric of space is homogeneous, therefore the elements of coupling matrix $G_{AB}$ and $\tilde{G}_{AB}$ is just function of $\eta$. 

The dynamics equations of $\mathbb{A}_a^A$ can be obtained  through \eqref{e23} and \eqref{e24}
\begin{equation}
	(\mathbb{A}_a^A)''+G^{AB}G'_{BC}(\mathbb{A}_a^C)'+G^{AB}\tilde{G}_{BC}'\epsilon^{bc}_{~~c}D_b\mathbb{A}_c^C-\delta^{bc}D_bD_c\mathbb{A}_a^A=0
	\label{e26}
\end{equation}
where $\epsilon_{abc}=a^{-3}\hat{\epsilon}_{abc}$ is Levi-Civita tensor and $'$ denote the derivative with respect to conformal time.

To get the power spectrum during the inflation era, the $\mathbb{A}_a^A$ should be expanded by plane wave:
\begin{equation}
	\label{e27}
	\mathbb{A}_a^A=\int\frac{d^3k}{(2\pi)^3}\sum_{h}\left(\bar{A}_h^Ab^he^{i\vec{k}\cdot\vec{x}}+\bar{A}^{A*}_hb^{h\dagger}e^{-i\vec{k}\cdot\vec{x}}\right)\delta_{ab}\hat{e}_h^b
\end{equation}
where $\hat{e}_h^a:=a e_h^a$ and $e_h^a$ is orthonormal polarization vectors and $h=\pm 1$ denote the different helicity.

$b^h,b^{h\dagger}$ are annihilation and creation operators which satisfy
\begin{equation}
	\label{e28}
	\langle0|b^{h}(\vec{k})b^{h'\dagger}(\vec{k}')|0\rangle=(2\pi)^3\delta^{hh'}\delta^3(\vec{k}-\vec{k}')
\end{equation}
The mode function $\bar{A}_h^A$ satisfy following evolution equations
\begin{equation}
	\label{e29}
	(\bar{A}^A_h)''+\mathbb{Z}^A_B(\bar{A}_h^B)'+hk\tilde{\mathbb{Z}}^A_B\bar{A}_h^B+k^2\bar{A}_h^A=0
\end{equation}
where
\begin{equation}
	\label{e37-a}
	\mathbb{Z}^A_B:=G^{AC}G'_{CB},~~~\tilde{\mathbb{Z}}^A_B:=G^{AC}\tilde{G}'_{CB}
\end{equation}

From \eqref{e29} one can see that the evolution of different vector fields will be coupled with each other. In this paper, we want to get the approximate analysis solutions, therefore, it is convenient to dispose of the first-time derivative terms and decouple the equations.
To achieve the purpose of decoupling the equations and removing the first-order terms, we set
\begin{equation}
	\label{e37}
	\mathcal{A}^A:=T^A_{~B}\bar{A}_h^B,~~\text{i.e.}~~\bar{A}_h^A=T^{~A}_B\mathcal{A}^B
\end{equation}
where $T^A_{~B}$ is a transformation matrix and $T^{~A}_B$ is its inverse matrix. Then the evolution equations for $\mathcal{A}^A$ are
\begin{equation}
	\label{e38}
	(\mathcal{A}^A)''+K^A_B(\mathcal{A}^B)'+\Omega^A_B\mathcal{A}^B+k^2\mathcal{A}^A=0
\end{equation}
where 
\begin{align}
	K^A_B:=&2T^A_{~C}(T_B^{~C})'+T^A_{~C}\mathbb{Z}^C_DT^{~D}_B\label{e39}\\
	\Omega^A_B:=&T^A_{~C}(T^{~C}_B)''+T^A_{~C}\mathbb{Z}^C_D(T_B^{~D})'+hkT^A_{~C}\tilde{\mathbb{Z}}^C_DT_B^{~D}\label{e40}
\end{align}
To remove the first-order term, the matrix $\mathbf{T}$ should satisfy
\begin{equation}
	\label{e41}
	K^A_B=0~\Rightarrow~(\mathbf{T}^{-1})'=-\frac{1}{2}\mathbf{\mathbb{Z}}\cdot\mathbf{T}^{-1}
\end{equation}
Insert \eqref{e41} into \eqref{e40} one can get:
\begin{equation}
	\label{e42}
	\mathbf{\Omega}=-\frac{1}{2}(\mathbf{T}\cdot\mathbf{\mathbb{Z}}\cdot\mathbf{T}^{-1})'-\frac{1}{4}(\mathbf{T}\cdot\mathbf{\mathbb{Z}}\cdot\mathbf{T}^{-1})^2+hk(\mathbf{T}\cdot\tilde{\mathbf{\mathbb{Z}}}\cdot\mathbf{T}^{-1})
\end{equation}
and the \eqref{e38} can be rewritten as
\begin{equation}
	\label{e43}
	(\mathcal{A}^A)''+\Omega^A_B\mathcal{A}^B+k^2\mathcal{A}^A=0
\end{equation}

From \eqref{e42}, one can see that, in$\mathbf{\mathbb{Z}}=\tilde{\mathbf{\mathbb{Z}}}$ case, the $\mathbf{\Omega}$ is diagonal matrix when the $\mathbf{T}$ matrix is formed by the linear independence eigenvectors of the $\mathbf{\mathbb{Z}}$ matrix, then \eqref{e43} is decouplead.
Therefore the key problem is the constructing of $\mathbf{T}$ matrix.

The $\mathbb{Z}$ matrix is determined by the coupling matrix $G_{AB}$, therefore $\mathbf{T}$ matrix which is formed by the linear independence eigenvectors of $\mathbb{Z}$ is also related to the coupling matrix. However, $\mathbf{T}$ cannot be completely determined by the coupling matrix. This is because the eigenvectors of $\mathbb{Z}$ which is used to form $\mathbf{T}$ are not necessary to be normalized, which means that it still has some arbitrariness when constructing the $\mathbf{T}$ matrix.

This arbitrariness of $\mathbf{T}$ can be eliminated by requiring $\mathbf{T}$ to meet Eq.\eqref{e41}. In fact, Eq.\eqref{e41} is overdetermined to get $\mathbf{T}$ matrix in some cases (see Sec.\ref{s4}), therefore it also limits the form of the coupling matrix $G_{AB}$.  

The energy density of the electromagnetic field can be get by \eqref{e11}
\begin{equation}\label{e30}
	\rho_{EM}=\frac{1}{8\pi}\langle 0\left|G_{AB}h^{ab}E_a^AE_b^B+G_{AB}h^{ab}B_a^AB_b^B\right|0\rangle=\rho_E+\rho_B
\end{equation}
where 
\begin{equation}\label{e31}
	\rho_E=\int \frac{dk}{k}\mathcal{P}_E,~~~~\rho_B=\int \frac{dk}{k}\mathcal{P}_B
\end{equation}
and the power spectrum is 
\begin{align}
	\mathcal{P}_E=&\sum_h\frac{G_{AB}}{16\pi^3}\frac{k^3}{a^4}(\bar{A}^A_h)'(\bar{A}^{B*}_h)'\label{e32}\\
	\mathcal{P}_B=&\sum_h\frac{G_{AB}}{16\pi^3}\frac{k^5}{a^4}\bar{A}^A_h\bar{A}^{B*}_h\label{e33}
\end{align}

Orthonormalization condition satisfied by the mode function can be obtained through isochronous commutation relationship between $\mathbb{A}_a^A,\Pi_B^b$
\begin{equation}
	\label{e34}
	\left[\mathbb{A}_a^A(\vec{x},\eta),\Pi_B^b(\vec{y},\eta)\right]=i\int\frac{d^3k}{(2\pi^3)}e^{i\vec{k}\cdot(\vec{x}-\vec{y})}\delta_B^A\left(\delta_a^b-\delta_{ac}\frac{k^bk^c}{k^2}\right)
\end{equation}
Insert \eqref{e27} into \eqref{e34} and use the commutation relationship \eqref{e28}, one can have
\begin{align}
	\sum_h\left[\bar{A}_h^A(\bar{A}^{B*}_h)'-\bar{A}^{A*}_h(\bar{A}_h^B)'\right]=&4\pi iG^{AB}\label{e35}\\
	\sum_h\left(\bar{A}^{A*}_h\bar{A}^B_h-\bar{A}^A_h\bar{A}^{B*}_h\right)=&0\label{e36}
\end{align}
\eqref{e35} is just the generalized Wronskian condition for multiple vector fields case. It is worth noting that \eqref{e36} will degenerate into the trivial equation in single vector fields case, however in 
multiple fields case, this condition is non-trivial.

\section{Double-field model\label{s4}}
In this section, we consider the double-field model. We set the coupling matrix as
\begin{equation}
	\label{e46}
	G_{AB}=\tilde{G}_{AB}=
	\begin{pmatrix}
		f^2_+(\eta)&~&I(\eta)\\
		~&~&~\\
		I(\eta)&~&f^2_-(\eta)
	\end{pmatrix}
\end{equation}
where $f^2_{\pm}$ denote the self-coupling of two vector fields and $I$ denote the mutual-coupling between two vector fields.
From \eqref{e37-a} one can get the $\mathbb{Z}$ matrix in double-field model as:
\begin{equation}
	\label{e47}
	\mathbb{Z}^A_B=G^{-1}
	\begin{pmatrix}
		2f_+f'_+f^2_--II'&~&f^2_-I'-2If_-f_-'\\
		~&~&~\\
		f^2_+I'-2If_+f_+'&~&2f^2_+f_-f_-'-II'
	\end{pmatrix}
\end{equation}
where $G=f^2_+f^2_--I^2$ is the determinant of the coupling matrix. We will discuss three non-helicity ($h=0$) cases below
\subsection{Case 1: no mutual-coupling case ($I=0$)}
A simple case of the double-field model is only considering the self-coupling of the vector fields without considering the mutual-coupling between them.

When $I=0$, $G_{AB}=diag\{f^2_+,f^2_-\}$, and $\mathbb{Z}^A_B=diag\{2f_+'/f_+,2f_-'f_-\}$. As mentioned before, the $\mathbf{T}$ matrix can be constructed by using eigenvectors of $\mathbb{Z}$. In fact, since $\mathbb{Z}$ is a diagonal matrix,  then any diagonal matrix can be used as $\mathbf{T}$ matrix. Therefore we set 
\begin{equation}
	\label{e48}
	\mathbf{T}^{-1}=
	\begin{pmatrix}
		g_+(\eta)&~&0\\~\\0&~&g_-(\eta)
	\end{pmatrix}
\end{equation}
where $g_+(\eta),g_-(\eta)$ can be decided by \eqref{e41} and this lead $g_{\pm}=f_{\pm}^{-1}$, which means $\mathbf{\Omega}=diag\{-f_+''/f_+,-f_-''/f_-\}$.
Now the evolution equations of mode function \eqref{e43} can be written as
\begin{equation}
	\label{e49}
	(\mathcal{A}^{\pm})''+\left(k^2-\frac{f_{\pm}''}{f_{\pm}}\right)\mathcal{A}^{\pm}=0
\end{equation}
At the beginning of inflation, all the scales of interest are all well sub-horizon, which means that $k\rightarrow\infty$. Therefore the initial conditions can be chosen as B-D vacuum
\begin{equation}\label{e50}
	\lim_{-k\eta\rightarrow \infty}\mathcal{A}^{\pm}\rightarrow \sqrt{\frac{2\pi}{k}}\exp(-ik\eta)
\end{equation}
It can be verified that these initial conditions satisfy \eqref{e35},\eqref{e36}. In this subsection, we consider the power law form of $f_{\pm}\propto\eta^{\gamma_{\pm}}$. 

The evolution equations of $\mathcal{A}^{\pm}$ are the same as in the single vector field model, i.g. \cite{subramanian2016origin,PhysRevD.96.083511}. However, in the double-field model, \eqref{e49} shows that the evolution of $\mathcal{A}^+$ will be different with $\mathcal{A}^-$ because of the difference between $f_+$ and $f_-$. This is the main difference between this case and the single vector field model.

The solutions of \eqref{e49} with the initial conditions \eqref{e50}are
\begin{equation}
	\label{e51}
	\mathcal{A}^{\pm}(k,\eta)=\pi\sqrt{-\eta}\left\{\exp\left[-i\frac{\pi\gamma_{\pm}}{2}\right]\frac{J_{\gamma_{\pm}-1/2}(-k\eta)}{\cos(\pi\gamma_{\pm})}+\exp\left[i\frac{\pi(\gamma_{\pm}+1)}{2}\right]\frac{J_{-\gamma_{\pm}+1/2}(-k\eta)}{\cos(\pi\gamma_{\pm})}\right\}
\end{equation}
where $J$ is Bessel functions. Insert this solution into \eqref{e32} and \eqref{e33} and take the $-k\eta\rightarrow 0$ limit one can get the behavior of power spectrum of the electromagnetic field at the end inflation:
\begin{align}
	\mathcal{P}_B\rightarrow&\mathcal{P}_{B+}+\mathcal{P}_{B-}=\frac{H^4}{8\pi^3}\left[\mathfrak{C}^2(n_+)(-k\eta)^{2n_++4}+\mathfrak{C}^2(n_-)(-k\eta)^{2n_-+4}\right]\label{e52}\\
	\mathcal{P}_E\rightarrow&\mathcal{P}_{E+}+\mathcal{P}_{E-}=\frac{H^4}{8\pi^3}\left[\mathfrak{D}^2(m_+)(-k\eta)^{2m_++6}+\mathfrak{D}^2(m_-)(-k\eta)^{2m_-+6}\right]\label{e53}
\end{align}
where 
\begin{equation}
	\label{e54}
	n_{\pm}:=
	\begin{cases}
		\gamma_{\pm}&(\gamma_{\pm}<1/2)\\
		1-\gamma_{\pm}&(\gamma_{\pm}>1/2)
	\end{cases}
	~~~~~~
	m_{\pm}:=
	\begin{cases}
		\gamma_{\pm}&(\gamma_{\pm}<-1/2)\\
		-1-\gamma_{\pm}&(\gamma_{\pm}>-1/2)
	\end{cases}
\end{equation}
and
\begin{align}
	\mathfrak{C}(x):=&\frac{\pi}{2^{x-1/2}}\frac{\exp(-i\pi x/2)}{\Gamma(x+1/2)\cos(\pi x)}\label{e55}\\
	\mathfrak{D}(x):=&\frac{\pi}{2^{x+1/2}}\frac{\exp(-i\pi x/2)}{\Gamma(x+3/2)\cos(\pi x)}\label{e56}
\end{align}
It can be seen that if one set $\gamma_+=2,\gamma_-=-2$, then $n_+=-1, n_-=-2, m_+=-3, m_-=-2$ and the power spectrum are
\begin{align}
	\mathcal{P}_B&\rightarrow\frac{H^4}{8\pi^3}\left[\mathfrak{C}^2(-1)(-k\eta)^2+\mathfrak{C}^2(-2)\right]\approx\frac{H^4}{8\pi^3}\mathfrak{C}^2(-2)\label{e57}\\
	\mathcal{P}_E&\rightarrow\frac{H^4}{8\pi^3}\left[\mathfrak{D}^2(-3)+\mathfrak{D}^2(-2)(-k\eta)^2\right]\approx\frac{H^4}{8\pi^3}\mathfrak{D}^2(-3)\label{e58}
\end{align}
It means that both electric and magnetic fields are scale-invariant spectrum when the inflation is finished. This conclusion is different from the single-field case, in which electric and magnetic fields spectrum cannot be scale-invariant at the same time. Therefore, the backreaction problem can be avoided in this case. 

On the other hand, in this case, $f_+\propto\eta^2$ is decreasing function, and $f_-\propto\eta^{-2}$ is increasing function.
If one set $f_+\sim1$ at the end of inflation and $f_-\sim 1$ at the beginning of inflation, then at the beginning of inflation, $f_+\gg f_-\sim1$, which means that the interaction between $A^+$ field and charged field is very weak, while the interaction between $A^-$ field and charged field is consistent with Maxwell's theory. 

At the end of inflation, $f_-\gg f_+\sim1$, the situation is just the opposite i.e. the interaction between $A^-$ field and charged field is very weak, while the interaction between $A^+$ field and charged field is consistent with Maxwell's theory.  
Meanwhile, during inflation, the self-coupling functions stay greater than one always, which means that there is no strong coupling problem. 

{
	From \eqref{e57} and \eqref{e58}, one can see that, the total magnetic field mainly comes from the contribution of $A^-$ and the total electric field comes from the contribution of $A^+$ at the end of inflation. In other words, the electric field and magnetic field after inflation do not come from the same vector field.}

{
After the inflation, the electrical conductivity of the cosmic plasma increased rapidly, which caused the electric field to degenerate to zero and the magnetic field expected to be observed mainly comes from $A^-$. Therefore, shortly after inflation ends, only the vector field $A^-$ remains and appears as a magnetic field, which means that there is no additional relativistic degree of freedom after inflation ends.  }

{
	Since $f_-$ is much larger than one, the magnetic field after the end of inflation is weakly coupled with the charged particle field. However, a vector field is a gauge field of a charged particle field, and multiple vector fields usually imply multiple charged particle fields as well.
}

{
	Assume that there are charged particle fields $\psi_{\pm}$ (e.g. fermion fields) corresponding to the $A_{\pm}$ fields. $f_-\gg1$ means the weakly coupled between $A^-$ and $\psi_-$. However, it does not mean the weakly coupled between $A^-$ and $\psi_+$. The models of interaction between different branches of different multiple fields have been discussed e.g. in \cite{babu_implications_1998,loinaz_charge_1999}. 
}

{
	Therefore, if it is assumed that the coupling between the $A^-$ field and the $\psi_+$ field is normal, then this magnetic field will interact with the post-inflation cosmic plasma  like an ordinary magnetic field, and thus become the primordial magnetic field today. On the other hand, the $\psi_-$ field will decay into the dark sector of the universe due to the weak coupling.
}

In summary, in case 1, the power spectrum of electric and magnetic fields are both scale-invariant at the end of inflation, meanwhile, there is no backreaction and strong coupling problem, and the interaction with charged fields is "normal" at the end of inflation.

\subsection{Case 2: weak mutual-coupling case ($I'\ll I\ll 1$ )}
In this subsection, we introduce a weak and nearly constant mutual-coupling $I$ ($I'\ll I\ll 1$ ) based on case 1. Therefore, the main conclusions will be the same as in case 1, and we only discuss the effect of weak coupling between two vector fields on 
these conclusions. We set $f_{\pm}\propto\eta^{\pm\gamma}$. To avoid the strong coupling problem, we also require $f_{\pm}>1$ as in case 1. Then in case 2
\begin{equation}
	\label{e59}
	\mathbb{Z}_A^B\approx
	\begin{pmatrix}
		2\frac{f_+'}{f_+}\left(1-\frac{I^2}{f^2_+f^2_-}\right)&~&-2I\frac{f_-'}{f^2_+f_-}\\
		~\\
		2I\frac{f_+'}{f_+f^2_-}&~&2\frac{f_-'}{f_-}\left(1-\frac{I^2}{f^2_+f^2_-}\right)
	\end{pmatrix}
\end{equation}
and the matrix $\mathbf{T}$ can be written as 
\begin{equation}
	\label{e60}
	\mathbf{T}^{-1}=
	\begin{pmatrix}
		g_+&~&-g_-\mathcal{I}_-\\
		~\\
		g_+\mathcal{I}_+&~&g_-
	\end{pmatrix}
\end{equation}
where 
\begin{equation}\label{e61}
	\mathcal{I}_{\pm}:=\frac{If_{\pm}'}{f_{\pm}f^2_{\mp}}\left(\frac{f_{\pm}'}{f_{\pm}}-\frac{f_{\mp}'}{f_{\mp}}\right)^{-1}\propto O(I)\ll 1
\end{equation}

Insert $\mathbf{T}^{-1}$ in \eqref{e41} one can found that there is no $g_{\pm}$ can satisfy \eqref{e41}  unless $I=0$. The reason is that \eqref{e41} is overdetermined in this case as we discuss in Sec.\ref{s3}. 

This means that if one still uses the eigenvectors of $\mathbb{Z}$ matrix to construct $\mathbf{T}$ matrix, then the coupling term between $\mathcal{A}^+$ and $\mathcal{A}^-$ and the first time derivative term will appear in the evolution equation \eqref{e38}.
The reason for these terms is the off-diagonal elements of $\mathbf{T}^{-1}$ which are order $O(I)$. From \eqref{e39} and \eqref{e40}, one can find that these terms are also order $O(I)$, which can be omitted compared to the other term. Therefore, we ignored the off-diagonal equations in \eqref{e41}.
Then \eqref{e41}  is 
\begin{equation}
	\label{e62}
	\begin{pmatrix}
		g_+'+g_+\frac{f_+'}{f_+}&~&g_+\mathcal{I}_+\left[\frac{(g_+\mathcal{I}_+)'}{g_+\mathcal{I}_+}+\frac{f_+'}{f_+}\right]\\
		~\\
		g_-\mathcal{I}_-\left[\frac{(g_-\mathcal{I}_-)'}{g_-\mathcal{I}_-}+\frac{f_-'}{f_-}\right]&~&g_-'+g_-\frac{f_-'}{f_-}
	\end{pmatrix}
	\approx
	\begin{pmatrix}
		g_+'+g_+\frac{f_+'}{f_+}&~&0\\
		~\\
		0&~&g_-'+g_-\frac{f_-'}{f_-}
	\end{pmatrix}
	=0
\end{equation}
which means that $g_{\pm}=f_{\pm}^{-1}$ as in case 1. Therefore $\mathbf{\Omega}=diag\{-\mu_+\eta^{-2},-\mu_-\eta^{-2}\}$ where
\begin{equation}
	\label{e63}
	\mu_{\pm}:=\gamma\left[\left(\gamma\mp 1\right)-\left(\gamma\mp \frac{1}{2}\right)I^2q^2\right]
\end{equation}
and $q:=(f_+f_-)^{-1}=\text{constant}$. \eqref{e43} now are
\begin{equation}
	\label{e64}
	(\mathcal{A}^{\pm})''+\left(k^2-\frac{\mu^{\pm}}{\eta^2}\right)\mathcal{A}^{\pm}=0
\end{equation}
and the power spectrum of magnetic fields at the end of inflation is 
\begin{equation}
	\label{e65}
	\mathcal{P}_B=\frac{H^4}{8\pi^3}\left[\mathfrak{C}^2\left(-\nu_+\right)(-k\eta)^{-2\nu_++5}+\mathfrak{C}^2\left(-\nu_-\right)(-k\eta)^{-2\nu_-+5}\right]
\end{equation}
where $\nu_{\pm}:=\frac{1}{2}\sqrt{1+4\mu_{\pm}}$ and we ignore the contributions of off-diagonal elements of $G_{AB}$ and $\mathbf{T}$ because their contributions to the power spectrum are $O(I^2)$.
Because $\mu_->\mu_+$ then $\nu_->\nu_+$, therefore the second term dominant the power spectrum
\begin{equation}
	\label{e66}
	\mathcal{P}_B\sim \frac{H^4}{8\pi^3}\mathfrak{C}^2\left(-\nu_-\right)(-k\eta)^{-2\nu_-+5}
\end{equation}
and the index of the spectrum is $n_B=-2\nu_-+5$. If one chose $\gamma=2$, then $n_B=2I^2q^2\propto O(I^2)$.

The above conclusion shows that the weak coupling between the two vector fields can make the scale-invariant magnetic field spectrum slightly tilt towards the blue and the index of the spectrum is proportional to the square of the mutual-coupling $I$ between the two vector fields.
\subsection{Case 3: strong mutual-coupling case ($f_+=1,f_-=f>1,I\gg f$)}
In this subsection, we consider the case in which the self-coupling $f_{\pm}=\text{constant}$. To avoid the strong coupling problem, we set$f_+=1,f_-=f>1$. 
Then 
\begin{equation}\label{e67}
	G_{AB}=
	\begin{pmatrix}
		1&~&I\\~\\
		I&~&f^2
	\end{pmatrix}
	~~\Rightarrow~~\mathbb{Z}_A^B=\frac{I'}{f^2-I^2}
	\begin{pmatrix}
		-I&~&f^2\\
		~\\
		1&~&-I
	\end{pmatrix}
\end{equation}
and
\begin{equation}
	\label{e68}
	\mathbf{T}^{-1}=
	\begin{pmatrix}
		-g_+f&~&g_-f\\~\\
		g_+&~&g_-
	\end{pmatrix}
\end{equation}
Insert \eqref{e67},\eqref{e68} into \eqref{e41} one can get 
\begin{equation}
	\label{e69}
	g_{\pm}=\frac{1}{\sqrt{I\mp f}}
\end{equation}
and 
\begin{equation}
	\label{e70}
	\mathbf{\Omega}=diag\left\{-\frac{I''}{2(I-f)}+\frac{1}{4}\left(\frac{I'}{I-f}\right)^2,-\frac{I''}{2(I+f)}+\frac{1}{4}\left(\frac{I'}{I+f}\right)^2\right\}
\end{equation}
The evolution equations \eqref{e43} now is 
\begin{align}
	(\mathcal{A}^+)''+\left[k^2-\frac{I''}{2(I-f)}+\frac{1}{4}\left(\frac{I'}{I-f}\right)^2\right]\mathcal{A}^+=&0\label{e71}\\
	(\mathcal{A}^-)''+\left[k^2-\frac{I''}{2(I+f)}+\frac{1}{4}\left(\frac{I'}{I+f}\right)^2\right]\mathcal{A}^-=&0\label{e72}
\end{align}
Considering $I\gg f$, then the evolution equations of $\mathcal{A}^{\pm}$ are the same. On the other hand, the initial conditions for $\mathcal{A}^{\pm}$ can also be both chosen as \eqref{e50} which satisfy \eqref{e35},\eqref{e36}. Therefore $\mathcal{A}^+\sim\mathcal{A}^-$ in case 3. The evolution equations of $\mathcal{A}^{\pm}$ are approximately
\begin{equation}
	\label{e73}
	(\mathcal{A}^{\pm})''+\left[k^2-\frac{1}{2}\frac{I''}{I}+\frac{1}{4}\left(\frac{I'}{I}\right)^2\right]
\end{equation}
We now consider the power law mutual-coupling $I\propto \eta^{\delta}$, then \eqref{e73} change to
\begin{equation}
	\label{e74}
	(\mathcal{A}^{\pm})''+(k^2-\alpha\eta^{-2})\mathcal{A}^{\pm}=0,~~~~\text{where}~~\alpha:=\frac{\delta^2}{4}-\frac{\delta}{2}
\end{equation}
The power spectrum of electromagnetic fields at the end of inflation is
\begin{align}
	\mathcal{P}_B\sim&\frac{H^4}{2\pi^3I_q}|\mathfrak{C}(-\beta)|^2(-k\eta)^{-2\beta+5}(-\eta)^{-\delta}\label{e75}\\
	\mathcal{P}_E\sim&\frac{H^4}{2\pi^3I_q}\left(-\frac{1}{2}\delta-\beta+\frac{1}{2}\right)^2|\mathfrak{C}(-\beta)|^2(-k\eta)^{-2\beta+3}(-\eta)^{-\delta}\label{e76}
\end{align}
where $\beta:=|\delta-1|/2$ and $I_q:=I_0(-\eta_f)^{-\delta}$ in which $I_0$ is the mutual-coupling at the beginning of inflation and $\eta_f$ is the conformal time at the end of inflation.

If the magnetic field is a scale-invariant spectrum, then
\begin{equation}
	\label{e77}
	n_B=-2\beta+5=0~~\Rightarrow~~\beta=\frac{5}{2}~~\Rightarrow~~\delta=6~\text{or}~\delta-4
\end{equation}
The electric field is a red spectrum because of $n_E=-2\beta+3=-2$. When $\delta=6$, $\mathcal{P}_B$ and $\mathcal{P}_E$ are both increasing with inflation and lead to backreaction problem. 
When $\delta=-4$, on the other hand, $\mathcal{P}_B$ and $\mathcal{P}_E$ are both decreasing with inflation and they all be diluted by inflation which means that there not enough primordial magnetic field generated.

While, if the electric field is a scale-invariant spectrum, then
\begin{equation}
	\label{e78}
	n_E=-2\beta+3=0~~\Rightarrow~~\beta=\frac{3}{2}~~\delta=4~\text{or}~\delta=-2
\end{equation}
The magnetic field is a blue spectrum because of $n_B=-2\beta+5=2$. When $\delta=4$, $\mathcal{P}_B$ and $\mathcal{P}_E$ are both increasing with inflation and lead to backreaction problem. 
When $\delta=-2$, on the other hand, $\mathcal{P}_B$ and $\mathcal{P}_E$ are both decreasing with inflation and they all be diluted by inflation which means there not be enough primordial magnetic field generated.

On the other hand, if one requires the spectrum of the magnetic field to be independent of time, then the value of $\delta$ should be $3$, and the electric field spectrum $\mathcal{P}_E\propto(-\eta)^{-2}$ will increase with inflation and lead to backreaction problem. In this case, $n_B=3$ and $n_E=1$ which means that they are both blue spectrum.

Similarly, if one requires the spectrum of the electric field is the independence of time, then the value of $\delta$ should be $2$, and the magnetic field spectrum $\mathcal{P}_B\propto (-\eta)^2$ will be diluted by inflation. In this case, $n_B=4$ and $n_E=2$ which means that they are both blue spectrum.

All in all, in case 3, if a sufficient primary magnetic field is to be generated, then the problem of backreaction is unavoidable.

\section{Summary and Discussion\label{s5}}
In this paper, we discuss the inflationary magnetogenesis base on the multiple vector fields model. The main problems of inflationary magnetogenesis are 
strong coupling broblem and backreaction problem which are both related to the coupling function $f(\phi)$. To avoid the strong coupling problem, $f(\phi)$ should be decreasing function. However, decreasing the coupling function will lead to a backreaction problem. Therefore, a single coupling function cannot solve both the strong coupling problem and the backreaction problem at the same time. 

One way to introduce more coupling functions is to consider the multiple vector fields. Instead of a single coupling function, we consider the coupling matrix whose matrix elements are a function of the scalar field $\phi$ which drive the inflation. We use the method of Hamiltonian dynamics to get the evolution equations of electromagnetic fields and the Gauss constraint first. Then apply these evolution equations to the inflation era.  

We found that in addition to the generalized Wornskian condition \eqref{e35}, there is another orthonormalization condition for mode function $\bar{A}^A_h$, i.e. \eqref{e36}. In the single vector model, this condition reduces to a trivial equation. However, in multiple vector fields cases, this condition is non-trivial. This condition limits the setting of initial conditions for the mode function as the Wornskian condition. In this paper, we set the initial condition of different mode functions to be the same BD vacuum \eqref{e50}, which satisfies both \eqref{e35} and \eqref{e36}.

Three types of the double-field model with no helicity have been considered in this paper. In case 1, we assume that there is no mutual interaction between the two fields and the self-interactions of these two vector fields with different coupling functions. The form of evolution equations for mode functions is the same as the single vector field case, however, the behavior of the electromagnetic field is different because of the difference in the self-coupling function. In this case, the spectrum of the electric field and magnetic field which are both scale-invariant can be obtained at the end of inflation. This is the main difference from the single vector field model in which electric and magnetic fields spectrum cannot be scale-invariant at the same time. Therefore the backreaction problem can be avoided in case 1. On the other hand,  the self-interaction coupling function stays greater than one during the inflation, therefore, the strong coupling problem can also be avoided and the interaction between the electromagnetic field and charged field is consistent with Maxwell's theory at the end of inflation.

Another factor affecting the electromagnetic field spectrum in the multiple vector fields model is the mutual-coupling between different vector fields. In case 2, we consider the weak mutual-coupling case base on case 1. Since the mutual interaction is weak, i.e. $I'\ll I\ll 1$, then the spectrum of the electric and magnetic field is still near scale-invariant at the end of inflation. The effect of weak mutual interaction between two vector fields is reflected in the spectrum index. 
In case 2, the spectrum index of the magnetic field is $n_B\approx 2I^2q^2\gtrsim 0$ which means that the mutual-coupling causes the energy density of the magnetic field to dissipate slightly with inflation.

To discuss the effects of mutual-coupling further, we also consider the strong mutual-coupling case. We found that a scale-invariant spectrum of magnetic is obtained in this case. However, when the spectrum of the magnetic field is scale-invariant, then the energy density of the electromagnetic field either increases with inflation and leads to the backreaction problem or is diluted by inflation which means there is not enough primordial magnetic field. On the other hand, if the spectrum of the magnetic field is time-independent, then the energy of the electric field will increase and lead to the backreaction problem.

In summary, in double-field models, if the self-coupling is constant, the mutual-coupling cannot generate enough primordial magnetic field and avoid the back-rection problem at the same time. Therefore the non-constant self-coupling is the necessary condition for double-field models of inflationary magnetogenesis. 

\section*{Acknowledgments}
This work was supported by the Fundamental Research Funds for the Central Universities of
Ministry of Education of China under Grants No. 3132018242,
the Natural Science Foundation of Liaoning Province of China under Grant No.20170520161 and the National Natural Science Foundation of China under Grant No.11447198 (Fund of theoretical physics).
\bibliographystyle{ws-mpla}
\bibliography{leeyu-mpla}
\end{document}